\documentclass[epjST]{svjour}
\usepackage{graphicx}
\usepackage{amsmath}
\usepackage{graphics,color}
\usepackage[square, numbers]{natbib}
\begin{document}
\title{Analysis of a bistable climate toy model with physics-based machine learning methods} 
\author{Maximilian Gelbrecht \inst{1,2}\fnmsep\thanks{\email{gelbrecht@pik-potsdam.de}} \and Valerio Lucarini\inst{3,4} \and Niklas Boers\inst{1,5,6} \and J\"urgen Kurths\inst{1,2,7}}
\institute{Potsdam Institute for Climate Impact Research, Potsdam, Germany \and Department of Physics, Humboldt-Universit\"at zu Berlin, Berlin, Germany \and Department of Mathematics and  Statistics, University of Reading,  Reading,  United  Kingdom \and Centre for the Mathematics of Planet Earth,  University of Reading, Reading, United  Kingdom \and Department of Mathematics and Computer Science, Freie Universit\"at Berlin, Berlin, Germany \and Department of Mathematics and Global Systems Institute, University of Exeter, Exeter, United Kingdom \and Lobachevsky State University of Nizhny Novgorod, Nizhny Novgorod, Russia}
\abstract{
We propose a comprehensive framework 
able to address both the predictability of the first and of the second kind for high-dimensional chaotic models. For this purpose, we analyse the properties of a newly introduced multistable climate toy model constructed by coupling the Lorenz '96 model with a zero-dimensional energy balance model. First, the attractors of the system are identified with Monte Carlo Basin Bifurcation Analysis. Additionally, we are able to detect  the Melancholia state separating the two attractors. Then, Neural Ordinary Differential Equations are applied in order to predict the future state of the system in both of the identified attractors.} 
\maketitle
\section{Introduction}
\label{intro}

Understanding the qualitative behaviour and dynamics of high-dimensional chaotic models of complex systems is a challenging task. The Earth's climate \citep{budyko69,sellers69,Lucarini_2020}, but also power grids \citep{machowski2008}, the human brain \citep{Babloyantz1986,Lytton2008}, and perception \citep{schwartz2012} or gene expression networks \cite{Smolen2000} all exhibit, in certain range of the system's parameters, multiple attractors with different basins of attractions. Energy balance climate models exhibit the well known hysteresis behaviour with respect to the solar radiation between a cold snowball earth state and a warmer state corresponding to the present-day climate, with discontinuous transitions taking place at the lower and upper boundary of the region of bistability \citep{budyko69,sellers69,Ghil1976,Lucarini2010,GhilLucarini2020,LucariniBodai2017,LucariniBodai2019PRL}. Recently, it has become apparent that the climate system might indeed feature more than two competing states, associated with a complex partitioning of the phase space in competing basins of attraction \cite{Lewis2007,Abbott2011,Brunetti2019,Margazoglou2020}. Indeed, multistability also almost always gives rise to a potential abrupt change of the system when a bifurcation point is - adiabatically - reached and one state loses it stability. In the context of geosciences, the crossing of a bifurcation point is usually referred to as tipping point \citep{Lenton1786}. Studying them is one of the crucial challenges of understanding and hopefully mitigating climate change \citep{lenton2019climate}. The notion of tipping point has been recently widened in order to accmomodate transitions that are caused specifically by the presence of a non-vanishing rate of change of the parameter of interest or by noise \cite{callaway2013dichotomy, Ashwin2012}. Far from the tipping point, when the system is a regime of multistability, transitions between the competing states is not possible in the case of autonomous dynamics, as the asymptotic state is determined by the initial condition, depending on which basin of attraction it belongs to. Initial conditions located on the basin boundaries - which have vanishing Lebesgue measure - are, instead, attracted to the edge states, which are saddles   located on such basin boundaries \cite{Grebogi1983,Vollmer2009,LucariniBodai2017}. Such saddles determine the global instabilities of the system and, additionally, if the system is, under fairly general conditions, forced with Gaussian noise, they are the gateways for noise-induced transitions between competing metastable states \cite{Graham1991,LucariniBodai2019PRL,Lucarini_2020,Margazoglou2020}. 

In order to characterise multistable systems, one needs to recover information on each competing attractor, so that it is possible to dynamically distinguish them. Indeed, one should be able to compute dynamical properties as e.g. different Lyapunov exponents or the basin stability \citep{Menck2013}. Any approach aimed at performing predictions in potentially multistable systems thus needs to first identify the different attractors and their basins of attractions to be able to make meaningful predictions on either of them. In this article, we will present a two-part framework to approach high-dimensional, spatiotemporally chaotic models, to understand their complex behaviour and their basins of attraction, and then predict future states of each of their attractors. In Lorenz' terminology, in this paper we try to bundle together a methodology to address both the predictability of the first kind - the model sensitivity to inaccurate initial conditions, hindering infinitely long deterministic predictions - and of the second  kind, associated with the uncertainty on the asymptotic state reached by the system \cite{Lorenz1975}.

We proceed as follows. The Monte Carlo Basin Bifurcation Analysis (MCBB) \citep{mcbb} is used to identify the basins of attractions with the largest volumes and how the volume changes when control parameters of the system are changed. MCBB makes use of random sampling and clustering techniques to quantify the volume of the basins of attraction and track how they change when control parameters of the system are varies. Based on the MCBB results, we learn which trajectories are asymptotically evolving towards which attractor.

For achieving predictability of the first kind, we apply a hybrid approach that complements potentially incomplete models with data-driven methods. Most numerical models describing a real world system and especially those of Earth system models, can be seen as incomplete in some regard. This can be due to e.g. neglecting higher order terms or by an unknown external influence. Predicting even only partially known chaotic systems has been successfully done with hybrid approaches that combine knowledge of the governing equation with data-driven methods \citep[e.g.]{pathak2018}. One particular promising approach are Neural Differential Equation or Universal Differential Equations \citep{chen2018neural,rackauckas2020universal} that enable us to train artificial neural networks (ANNs) inside of the differential equations.

We will test our framework on a new prototypical bistable model that we introduce below. The model is constructed by coupling a zero-dimensional classic energy balance model featuring bistability with the Lorenz96 (L96) model \citep{lorenz96}, which has gained prominence as prototypical system featuring spatially extended chaos. This model will serve as an ideal testbed for our approach.

The paper is structured as follows: First, we will introduce the Bistable Climate Toy Model, recap the dynamic properties of its key ingredient, the Lorenz96 model, and investigate the basic properties of the full coupled toy model. Subsequently, we will apply our two-part approach by first identifying and tracking the attractors of the model with MCBB, and then demonstrating the Neural Differential Equation approach on the model by replacing its energy balance model with an ANN and making predictions of the model. 

\section{Bistable Climate Toy Model} 

The Bistable Climate Toy Model is set up by coupling a Lorenz96 (L96) model \citep{lorenz96,Lorenz2005} to a zero-dimensional energy balance model (EBM). The L96 model describes a highly nontrivial dynamics on a one-dimensional periodic lattice composed of $N$ grid points, and is rapidly becoming a reference for studying non-equilibrium steady states in spatially extended systems. The L96 model features processes of advection, forcing, and dissipation. It can be thought of as representative of the dynamics of the atmosphere along one latitudinal circle \citep{lorenz96,Lorenz2005}, even if such correspondence is more metaphorical than actual, because the L96 model does not correspond to a truncated version of any known fluid dynamical system. The L96 model has rapidly gained relevance in many different research areas in order to study bifurcations \cite{vanKekem2018PhysD,vanKekem2018NPG}, to test parametrizations \cite{Wilks2005,Arnold2013,Vissio2018,Chattopadhyay2020}, to investigate extreme events \cite{Blender2013,Sterk2017,Hu2019}, to improve data assimilation schemes \cite{Trevisan2004,Brajard2020} and ensemble forecasting techniques \cite{Wilks2006,Duan2016}, to develop new tools for investigating predictability \cite{Hallerberg2010,Carlu2019}, and for addressing basic issues in non-equilibrium statistical mechanics \cite{AbramovM2008,Lucarini2011,Lucarini2012,Gallavotti2014}. The L96 model is formulated as follows:
\begin{equation}
\dot{X}_n = \left(X_{n+1}-X_{n-2}\right)X_{n-1}-\gamma X_n+F, \quad n=1,\ldots,N
\end{equation}
with periodic boundary conditions given by $X_{j-N}=X_j=X_{j+N}$ $\forall j=1,\ldots,N$. The parameter $F$ describes the forcing acting on the model, $\gamma$  controls the intensity of the dissipation and, thus, of the contraction of phase space volume, and the nonlinear term on the right hand side describes a non-standard advection. The energy of the system $E=1/2\sum_{n=1}^N X_n^2$ is conserved in the unviscid and unforced limit and acts as generator of the time translation, even though the system is not Hamiltonian. For a detailed analysis of the mechanics and energetics of the L96 model and of a generalisation thereof we refer to \cite{Vissio2020}.

If $\gamma=1$ (which is the default choice in most studies) and $N\gg1$, the model's attractor is the fixed point $X_k=F$, $k=1,\ldots,N$ for $0\leq F\leq8/9$. This fixed point loses stability as $F$ is increased and, after a complex set of bifurcations \cite{vanKekem2018PhysD,vanKekem2018NPG,vanKekem2019}, the system settles in a chaotic regime for$F\geq 5.0$ \cite{Lorenz2005}. In the regime of strong forcing and developed turbulence the properties of the L96 model are extensive with respect to the number of nodes $N$ \cite{Gallavotti2014,Vissio2020}, and one can establish power laws that accurately describe how some fundamental properties of the system - such as its energy and Lyapunov exponents - depend of the parameter $F$ \cite{Gallavotti2014}. 

As far as numerical evidence goes, in the turbulent  regime the L96 model possesses a unique asymptotic state characterised by a physical measure supported on a compact attractor. In order to introduce multistability in the L96 model, an efficient strategy is to suitably couple it with a multistable (say, bistable) model, according to the strategy described in \cite{Bodai2020}. Our simple bistable model of reference is the EBM of the form:
\begin{equation}
    \dot T=-\frac{\mathrm{d}V(T)}{\mathrm{d}T}
    \end{equation}
    where $V(T)$ is a confining potential ($V(T)\rightarrow \infty$ sufficiently fast as $|T|\rightarrow \infty$) with two local minima separated by a local maximum. We then come to the following coupled L96-EBM model:
\begin{align}
    \dot{T} &=  S\left( 1 - a_0 + \frac{a_1}{2}\left(\tanh\left(T -
    \tilde{T}\right)\right)\right) - \sigma T^4 - \alpha\left(\frac{\mathcal{E}(\mathbf{X})}{0.6\cdot F^\frac{4}{3}} -1\right)\nonumber\\
    \dot{X}_n &= \left(X_{n+1}-X_{n-2}\right)X_{n-1}-X_n+ F\left(1+\beta\frac{T-\tilde{T}}{\Delta_T}\right)\label{eq:model}
\end{align}
where  the usual periodic boundary conditions apply ($X_{j-N}=X_j=X_{j+N}$ $\forall j=1,\ldots,N$), and $\mathcal{E}=E/N$.  The value used for the different parameters - all intended to be non-negative - can be found in Tab. \ref{tab:parameters}. The L96 model and the EBM are uncoupled if one sets $\alpha=\beta=0$. The coupling between the two models can be explained as follows. If the temperature of the EBM is higher (lower) than the reference temperature $\tilde T$, the L96 model receives an enhanced (reduced) forcing, mimicking - in very rough terms - the presence of higher energy in the atmosphere. A negative feedback in the system is introduced as follows. If the energy per site of the L96 component of the model exceeds the average value realised in the uncoupled case $\bar{\mathcal{E}}\approx 0.6 F^{4/3}$ \cite{Gallavotti2014}, the temperature of the system is accordingly reduced. Note that, according to the framework set in \cite{Bodai2020}, the L96 model is the fast component and the EBM is the slow component of the coupled model, where the fast component acts as an almost stochastic forcing on the slow component, and the slow component modulates the dynamics of the fast component. We remark that since the coupling constants $\alpha$ and $\beta$ are $\mathcal{O}(1)$, one cannot use asymptotic methods such as averaging or homogenization to obtain a reduced equation for the temperature $T$; the dynamics of the system is truly high dimensional.

The system possess (at least) two competing attractors, associated with disjoint basins of attraction. Hence, the asymptotic state of the system depends on its initial conditions. We assume that each attractor possesses one physical measure that is practically selected when performing the numerical integration of the model. In absence of stochastic forcing, no transitions can take place between the two competing attractors.

\begin{figure}
    \centering
    \includegraphics[width=0.8\textwidth]{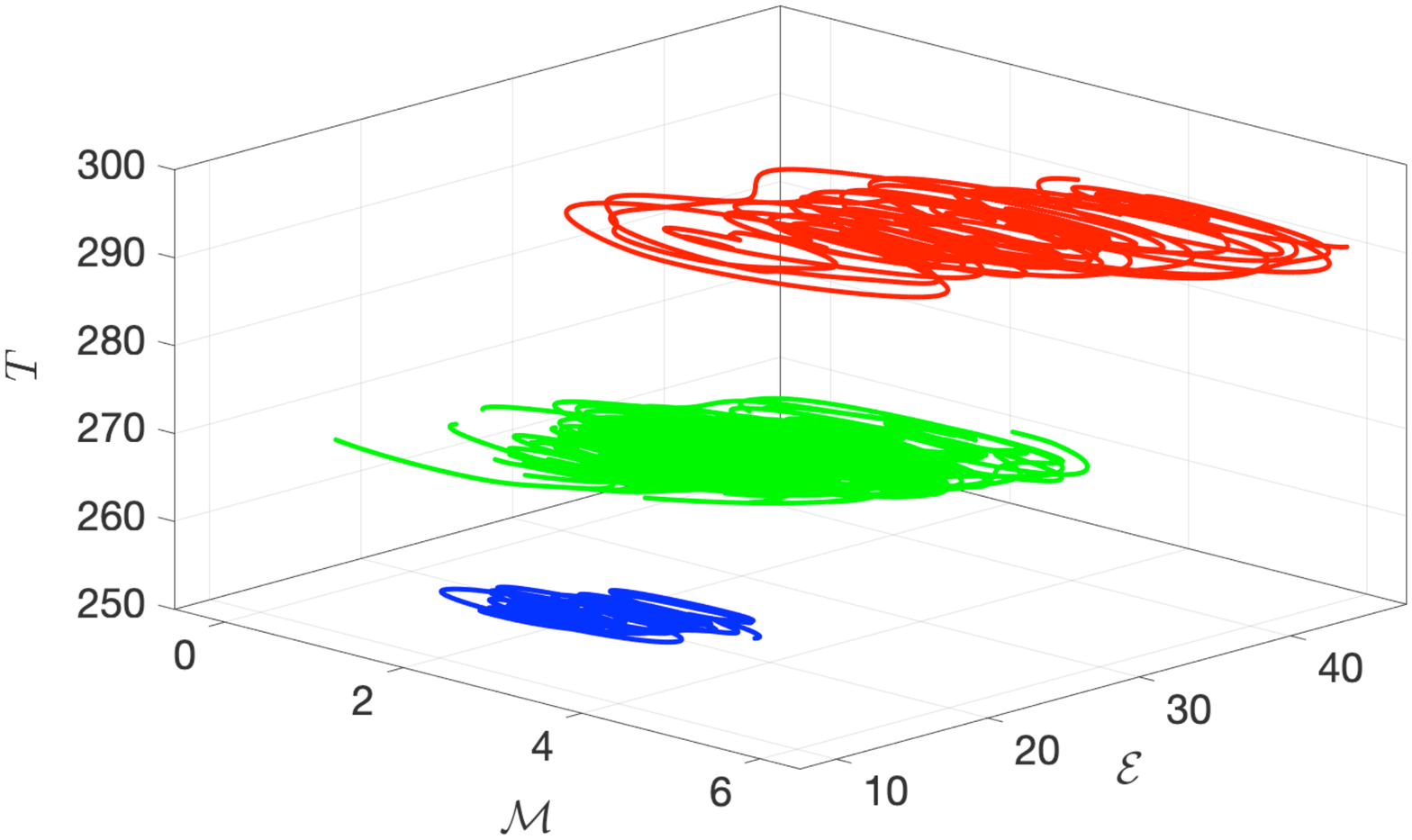}
    \caption{Competing Attractors and M state between them in the projection of the phase space given by $\mathcal{M}=1/N\sum_{j=1}^NX_j$, $\mathcal{E}=E/N$, and $T$. Simulations performed with $S=16$ and $N=32$. Warm (W): red line. Snowball (SB) state: blue line. Melancholia (M) State: green line.}
    \label{fig:mstate}
\end{figure}

In Fig.~\ref{fig:mstate} we portray the two competing attractors of the model corresponding to the Warm (W) state and Snowball (SB) state state in the reduced phase state constructed by performing a projection on the variables $\mathcal{M}=1/N\sum_{j=1}^NX_j$, $\mathcal{E}$, and $T$. One sees clearly that the W state has higher temperature and larger values for the mean and for the intensity of the fluctuations of the dynamic variables with respect to the SB state, in agreement with the actual features of the competing W and SB states of the climate system \cite{Pierrehumbert2011,Lucarini2010,Lucarini_2020}. We additionally portray the Melancholia (M) State sitting between the two competing attractors. The M State is a saddle embedded in the boundary between the two co-existing basins of attraction and attracts the orbits whose initial conditions are on such basin boundary \cite{LucariniBodai2017}. The M state, which is the gateway for the noise-induced transitions between the co-existing attractors regardless of the kind of noise included in the system \cite{LucariniBodai2019PRL,Lucarini_2020}, has been constructed using the edge-tracking algorithm \cite{Skufca2006}, and features non-trivial dynamics. Indeed, as in \cite{LucariniBodai2017}, it is a chaotic saddle.

\begin{table}[]
    \centering
    \caption{Parameters}

    \begin{tabular}{ccc}
      Reduced Solar Constant & $S$ & $5$ to $20$\\
      Albedo & $a_0$ & $0.5$ \\
      Albedo & $a_1$ & $0.4$ \\
      Reduced Stefan-Boltzmann Constant  & $\sigma$ & $1/180^4$\\ 
      Reference Forcing  & $F$ & $8$\\
      Number of Grid Points & $N$ & $40$ \\ 
      Reference Temperature & $\tilde{T}$ & 270\\
      Temperature Scale & $\Delta_T$ & 60 \\
      Coupling $X\rightarrow T$ & $\alpha$ & 2\\
      Coupling $T\rightarrow X$ & $\beta$ & 1 
    \end{tabular}
    \label{tab:parameters}
\end{table}

\section{Methods}

In the following, we outline a two-part framework to analyse and predict spatiotemporally chaotic system such as the Bistable Climate Toy Model. First we demonstrate how the attractors of the system can be can be identified. Based on that knowledge, a method to make predictions of states on both attractors is then introduced.

\subsection{Monte Carlo Basin Bifurcation Analysis} 

\begin{figure}
    \centering
    \includegraphics[width=0.5\textwidth]{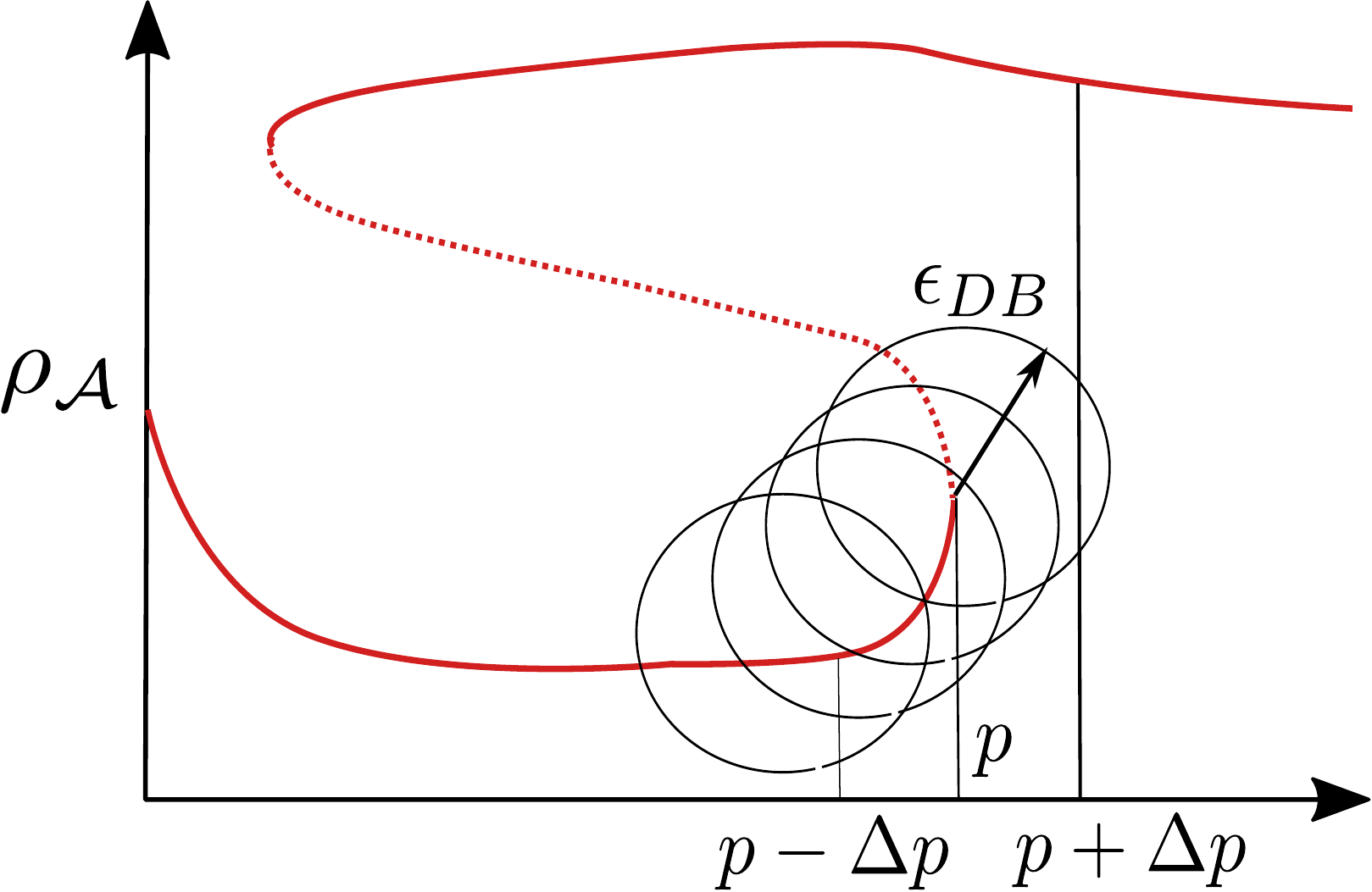}
    \caption{Sketch of the notion of continuity of invariant measures $\rho_\mathcal{A}$ that MCBB requires for attractors to belong to the same class of attractors. Solid red lines indicate stable states and dashed lines unstable states. If the difference between $\rho_\mathcal{A}(p)$ and $\rho_\mathcal{A}(p\pm\Delta p)$ vanishes smoothly for $\Delta\rightarrow 0$ they are considered to be within the same class. In the algorithm this is realised through a chain of connection via $\epsilon_{DB}$ neighbourhoods.}
    \label{fig:epsilon}
\end{figure}

Multistability is a universal phenomenon of complex systems and is most likely present in several sub-systems of Earth's climate \cite[e.g.]{Hirota232, Ciemer2019, May1977}, as well as in the energy balance of the Earth  \cite{budyko69,sellers69,Lucarini_2020}. When analysing and working with high-dimensional models, knowledge of the largest basins of attractions is instrumental to understanding the model itself. The recently introduced Monte Carlo Basin Bifurcation Analysis (MCBB) \citep{mcbb} is a numerical method tailored for analysing basins of attraction of high-dimensional systems and how they vary when control parameters of the system are changed. The aim of MCBB is to find classes of similar attractors of a high-dimensional system that collectively have the largest basin of attraction with respect to a measure of initial conditions $\rho_0$ and how these classes of attractors and their basin volumes change when a control parameter $p$ is changed in a range $I_p$. Conceptually, it is situated in between a thorough bifurcation analysis and a macroscopic order parameter. By utilizing random sampling and clustering techniques, MCBB learns classes of similar attractors $\mathcal{C}$ and their basins. To regard two attractors $\mathcal{A}$ as part of the same class, MCBB requires a notion of continuity of an invariant measure $\rho_\mathcal{A}$ on the attractor: if for a control parameter $p$ the difference between $\rho_\mathcal{A}(p)$ and  $\rho_\mathcal{A}(p+\Delta p)$ vanishes smoothly for $\Delta p\rightarrow 0$, we classify them as similar. Fig.~\ref{fig:epsilon} illustrates this continuity requirement. Note that if the change in the measure scales linearly with $\Delta p$ in the limit of small values of $\Delta p$, one can say that linear response theory applies for the measure $\rho_\mathcal{A}$ \cite{Ruelle2009}. By sampling trajectories these classes can be built with a suitable pseudometric on the space of these measures. Directly comparing the high-dimensional trajectories with each other would put us close to a bifurcation analysis, but is potentially prohibitively expensive. We might also not be interested in an in-depth bifurcation analysis, but rather in a coarser definition of similarity. For the case of a climate model, similar climate regimes may for example be of interest. 

Therefore, in order to identify the different classes of attractors, suitable statistics $\mathcal{S}_{i}$ are measured on every system dimension $k$ for every trajectory, here, the mean $E_k$, the variance $\text{Var}_k$ and the Kullback-Leibler divergence to a normal distribution $\text{KL}_k$. By using these statistics on each of the system dimension separately, MCBB achieves better scalability with the system dimension $N$. The $N$-dimensional trajectory of the $i$-th of in total $N_{T}$ trials is $\mathbf{x}^{(i)}(t)$, then all statistics are measured separately on each dimension of $\mathbf{x}^{(i)}(t)$. This results in three $(N\times N_{T}$) matrices, one for each statistic $\mathcal{S}_{i}$, so that $\mathbf{S}_{k,ij} = \mathcal{S}_k(x_j^{(i)})$. 

A distance matrix of each trajectory to each other is then computed from these statistics with 
\begin{align}
     D_{ij} = \sum_k^{3} w_k \sum_{l}^{N} |\mathbf{S}_{k,il} - \mathbf{S}_{k,jl}| + w_{4} | p^{(i)} - p^{(j)}|\label{eq:dist-direct}
\end{align}

where $p^{(i)}$ is the control parameter used to generate the $i$-th trajectory and $w_i$ are free parameters of the method.

If one wishes not to distinguish between symmetric configurations, as we also do in the application to the Bistable Climate Toy Model, the weighted difference can also be replaced with a Wasserstein distance of histograms over the statistics. Especially in these cases two different points $(x,y)$ may have $D(x,y) = 0$, this is why we are referring to $D$ as a pseudometric and not a metric. 
 
After computing the distance matrix $\mathbf{D}$ of all trials to each other, the classes of attractors can then be identified by applying a clustering algorithm to the matrix. Density-based clustering algorithms such as DBSCAN  \citep{Ester1996} are ideally suited for this purpose, since DBSCAN relies on a similar notion of continuity as required by MCBB. One sample is connected to another sample if it is within $\epsilon_{DB}$-neighbourhood of the sample. A cluster is then formed by all the samples that have a chain of connections to each other. The results of applying DBSCAN to sample trajectories is $\mathbf{C} = \text{DBSCAN}(\left\{\mathbf{D}\right\})$ where each sample trajectory is assigned to one of the $N_C$ clusters $C_i$ with $i\in[1;N_C]$. Note that DBSCAN can also identify samples as outliers. We regard all outliers, if there are any, as the zeroth cluster. The approximate relative basin size of a class of asymptotic states can then be computed by applying a sliding parameter window and normalizing the results with
\begin{align}
	\hat b_{\mathcal{C}_i}(p) &= ||{CL}_i^{(p)}|| / \sum_j^{N_C} ||CL_j^{(p)}||\\
	CL_i^{(p)} &= \left\{j | (C_j=i)\cap\left( p^{(j)}\in[p_{min};p_{max}]\right)\right\}.
\end{align}
Here, $CL_i^{(p)}$ is the number of trials in cluster $i$ at (sliding) parameter window $p$. This is the most important result from MCBB. $b_{\mathcal{C}_i}(p)$ shows the size of the largest basins of the system and how they react to changes of the control parameter $p$. In the results we will also see other possibilities to further evaluate the collected results from MCBB. By identifying the attractors, we also know which initial conditions are part of which basin. After achieving information about the attractors of the systems and their basins, we can investigate these individual attractor and apply further methods to predict trajectories on them. 

The MCBB method was designed with high dimensional systems in mind.  This is why per-dimension statistics are used as inputs for the clustering algorithm instead of the potentially high-dimensional trajectories directly. Nevertheless, the computational complexity of MCBB very  much depends on the system in question. The most expensive parts are $N$ times integrating the system and the computation of the distance matrix. For high-dimensional systems the integration far outweighs the computation of the distance matrix. For more details of MCBB the reader is referred to \cite{mcbb}. 

\subsection{Neural Ordinary Differential Equations} 

Most - if not all - numerical models of real-world processes are incomplete in some sense. Be it due to unknown effects that are not modelled, or by omitting higher-order terms of known effects on purpose. A classical example are subgrid-scale processes that are not explicitly resolved in general circulation models of the Earth's atmosphere and oceans, and hence need to be parametrized \citep{Berner2017,Franzke2015}. Hybrid modelling methods can remedy deficiencies of incomplete models by combining an incomplete process-based model with data-driven methods that learn to compensate these deficiencies. This approach seems especially promising with climate models as we have easy access to observational data and the climate system is so immensely complex that every model of it is always incomplete in some sense. We will demonstrate the Neural Ordinary Differential Equation (NODE) approach \citep{rackauckas2020universal,chen2018neural}, which provides an elegant way to construct and parametrize such hybrid models, on the Bistable Climate Toy model. 
NODEs integrate universal functions approximators such as artifical neural networks (ANN) directly into differential equation, so that the universal functions approximators become a part of the equation: 
\begin{align}
    \dot{x} = f(x,t,\mathcal{N}(x,t;\Theta)),
\end{align}
where $\mathcal{N}(x,t;\Theta)$ is a data-driven function approximator such as an Artificial Neural Network (ANN) with parameters $\Theta$. Integrating the NODE, just as integrating a regular ODE, yields a predicted trajectory $\hat{x}(t;\Theta)$ at discretized time steps $i_t$. The integration time can be freely set, but our previous results show that for chaotic systems very short integration times are necessary to ensure that the learning process succeeds \citep{npde-pre}. Similar to regular ANNs, NODEs are a supervised learning method and their parameters, in our case only the parameters of the ANN, are found by minimizing a loss function, most commonly the least-squares error between observed and model-predicted trajectories
\begin{align}
    \mathcal{L}(\Theta)=\sum_{i_t}\left( x(i_t) - \hat{x}(i_t;\Theta)\right)^2 \label{eq:L}  
\end{align}
using a stochastic, adaptive gradient descent with weight decay~\cite{loshchilov2017decoupled}. Additional regularization of the parameters of the ANN may be added to avoid overfitting. In order to train the model one needs to be able to compute gradients of the loss function with respect to all the parameters of the NODE. 

While a regular ANN relies on backpropagation -- which is essentially the chain rule -- to compute gradients of numerical solutions of differential equations is not as straight forward. However, with methods from adjoint sensitivity analysis and response theory in conjunction with automatic differentiation techniques, such gradients can be computed as well. For a detailed overview of the algorithms used for this purpose, please see \cite{rackauckas2020universal,chen2018neural}. 

\section{Results} 

\subsection{Attractors of the Model}

\begin{figure}
    \centering
    \includegraphics[width=0.6\textwidth]{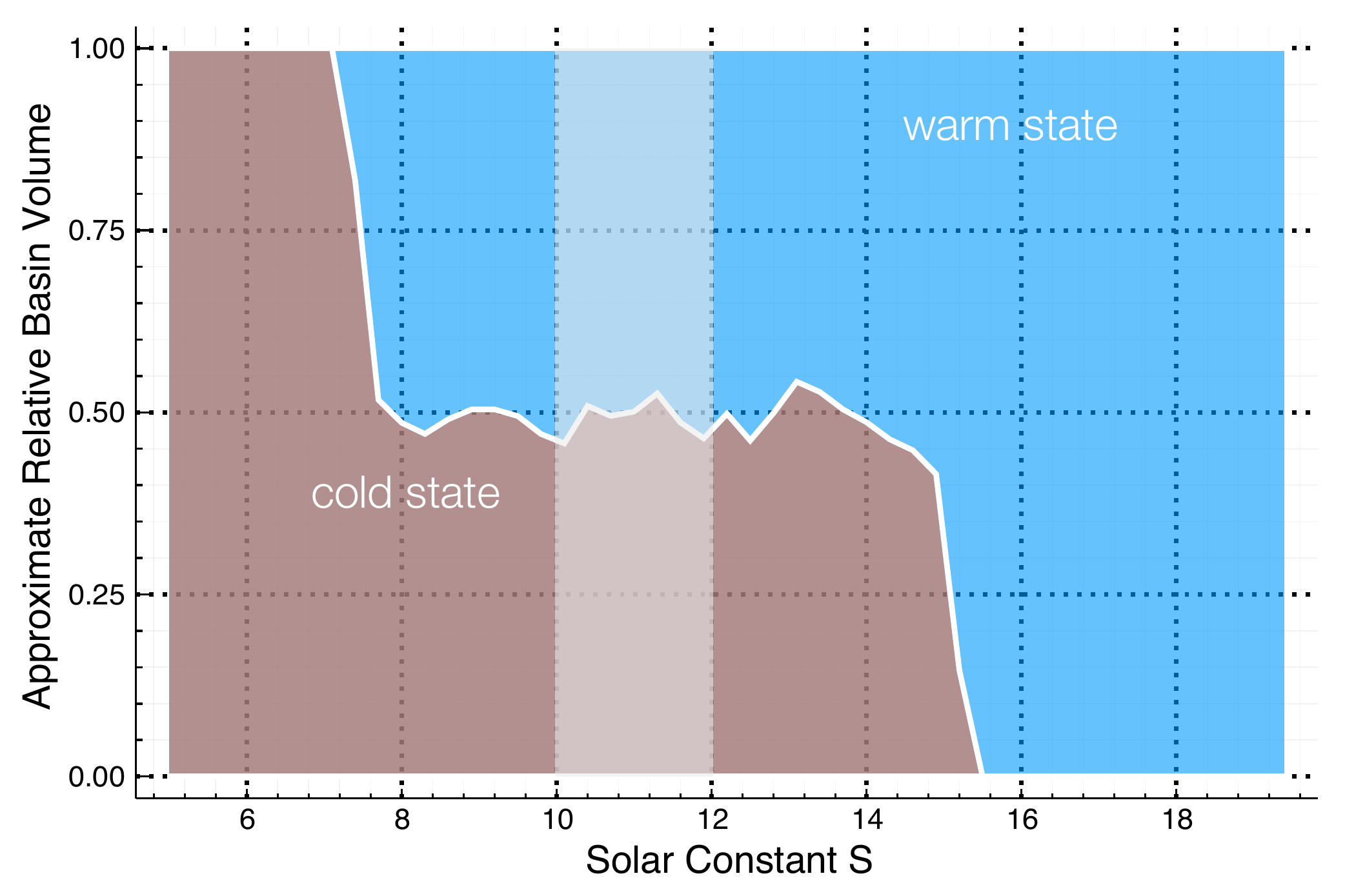}
    \caption{Approximate relative basin size of the Bistable Climate Toy Model when changing the solar constant $S$ estimated with MCBB. The model exhibits a cold and a warm state. Trajectories from the shaded area are used as training data for the NODE approach to predict each of these states.}
    \label{fig:mcbb-1}
\end{figure}

\begin{figure}
    \centering
    \includegraphics[width=0.7\textwidth]{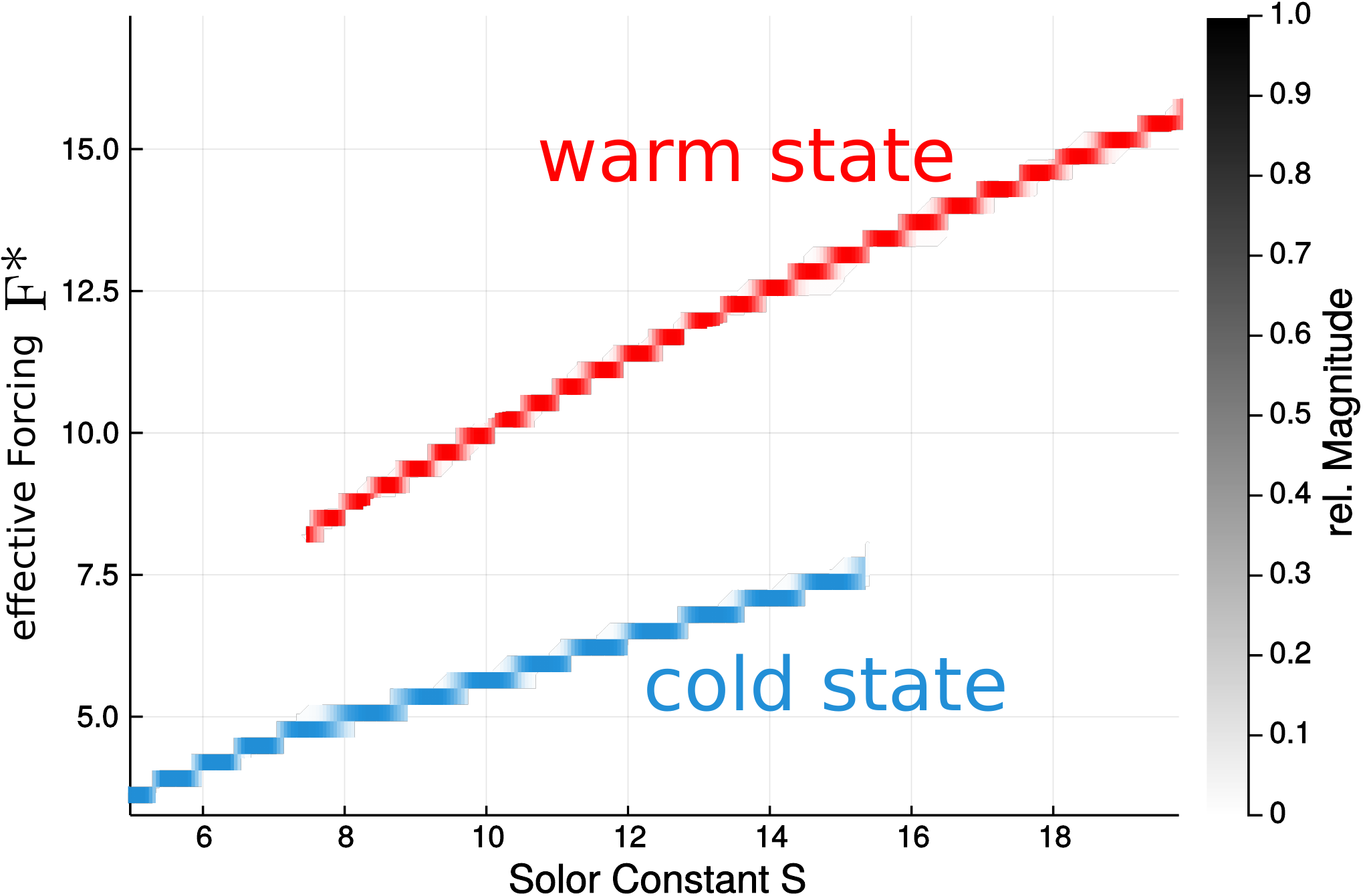}
    \caption{Sliding histogram plot of the mean of the EBM variable of the Bistable Climate Toy Model computed with MCBB. The two histogram plots of each of the identified states are joined together to better illustrate the two stable branches of the EBM and their hysteresis behaviour. On the y-axis the value of complete effective forcing term of the L96, so  $F*=  F\left(1+\beta\frac{T-\tilde{T}}{\Delta_T}\right)$, is shown. The relative magnitude of each of these values appearing in the individual sliding histograms is shown in shades of blue and red, however in this case are mostly all zero or one.} 
    \label{fig:mcbb-hist}
\end{figure}

\begin{figure}
    \centering
    \includegraphics[width=0.5\textwidth]{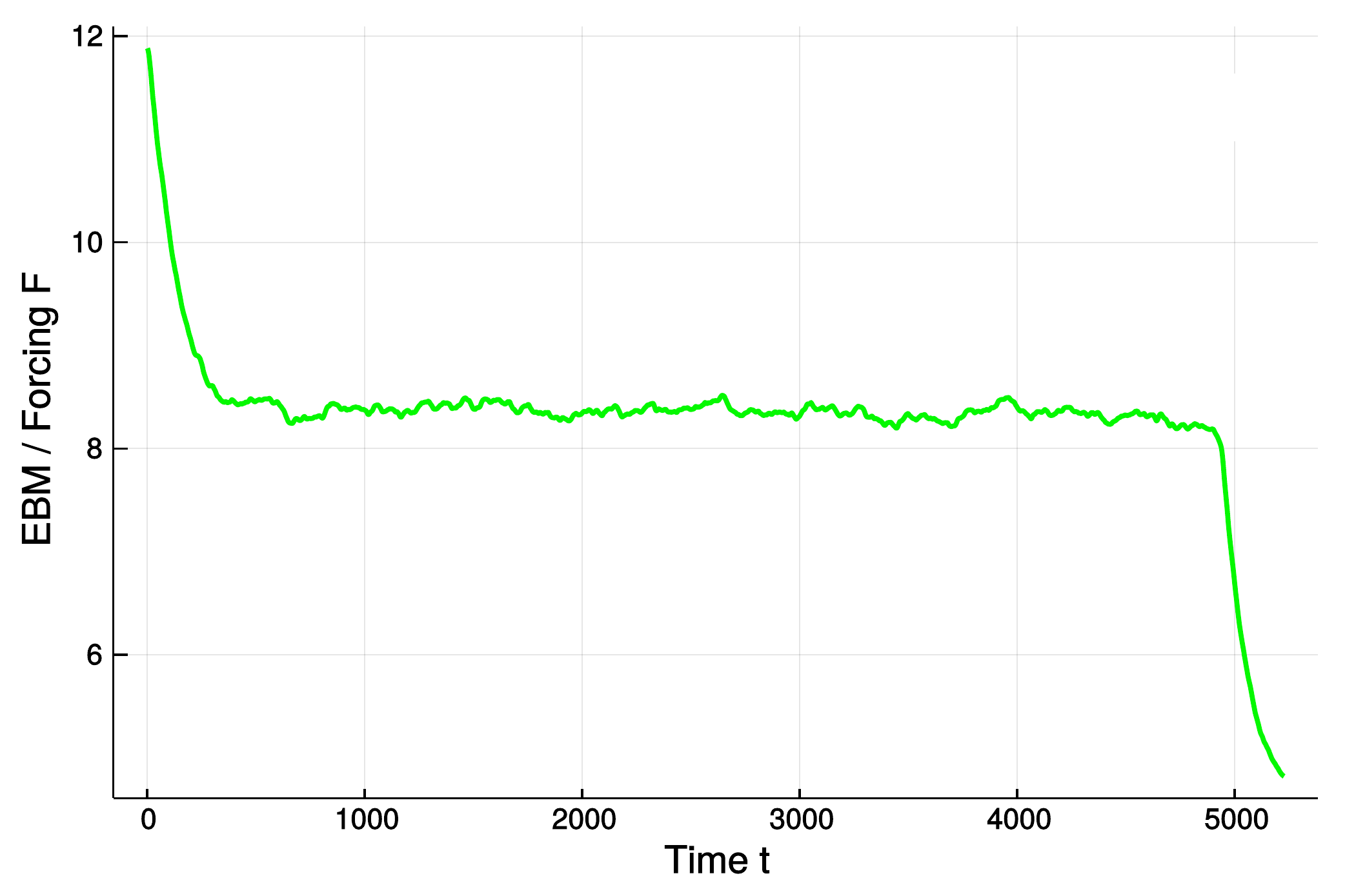}
    \caption{Example of the trajectory of the energy balance model for one of the Melancholia states found by MCBB as an outlier. Visible is a trajectory typical for a saddle. The trajectory remains close to the Melancholia state for some finite time until it collapses into the snowball/cold state.}
    \label{fig:m-example}
\end{figure}

We apply MCBB to the Bistable Climate Toy Model by sampling $N_T=15,000$ trajectories with initial conditions drawn from $\mathcal{U}(-7;7)$ for the L96 state variables $X$ and $\mathcal{U}(240,300)$ for the temperature $T$. The solar constant is varied within $S\in[5;20]$. The trajectories are integrated for $400$ time units and the first $80\%$ of the trajectory are not included in the analysis to avoid transient effects. Only the statistics $E_k$,$\text{Var}_k$,$\text{KL}_k$ of the L96 model are used for the identification of the attractors. In the distance matrix computation according to Eq.~\ref{eq:dist-direct}, the defaults weights of $[1,0.5,0.25]$ are chosen. The results are not sensitive to small variations of those weights. Fig.~\ref{fig:mcbb-1} shows the approximate relative basin volume estimated by MCBB. Two classes, i.e. clusters, are found. The system is multistable in the interval of around $S\in[7;15]$ with each of the basins being approximately equal-sized with respect to the distributions of initial conditions chosen. For $\epsilon_{DB}=0.05$ the clustering algorithm detects several outliers. These are mostly the trajectories that spend long time near the Melancholia state (M) because they are initialised near the basin boundary \cite{Lucarini2017}. As shown in Fig.~\ref{fig:m-example} they exhibit a saddle-like behaviour for the EBM variable. When $\epsilon_{DB}$ is increased to $0.1$ or larger these states are not resolved separately anymore and the algorithm only finds the cold and warm state as shown Fig.\ref{fig:mcbb-1}. Further insights can be gained with a sliding-histogram approach. For each sliding parameter window a histogram is fitted to all collected values of the EBM mean for each of the two attractors. Fig.~\ref{fig:mcbb-hist} clearly shows the two stable branches of the EBM and its respective values of the forcing $F$. As one can expect the "blue" cluster in Fig.~\ref{fig:mcbb-1} exhibits the much larger values of the forcing (see Fig. \ref{fig:mcbb-hist}b)) and is thus the warm state of the model and the "red" cluster in Fig.~\ref{fig:mcbb-1} is the cold state of the model. Again, the hysteresis behaviour of the model is evidently shown. For $S<7$ only the the cold state is stable and for $S>15$ only the warm state is stable.  

\subsection{Predicting the Model}

MCBB is thus able to identify the two attractors of the system. These two attractors will, in general, exhibit different properties, like e.g. different maximum Lyapunov exponent. The maximum Lyapunov exponents computed with the method of \citep{benettin1980} (implementation of DynamicalSystems.jl \citep{Datseris2018}) are $\lambda^{(\text{cold}}_{max}\approx1.04$ for the cold and $\lambda^{(\text{warm}}_{max}\approx2.60$ for the warm state. The larger Lyapunov exponent of the warm state shows that, as expected, the L96 model is more chaotic for larger values of the forcing. When we want to predict the model's behavior, we have to be aware of that and evaluate predictions on both attractors separately. MCBB also classifies all initial conditions that were used by the algorithms to either of the attractors. In that way, we have many possible initial conditions for prediction on these attractors. 
We demonstrate the capabilities of the NODE approach by "forgetting" the equation of the EBM and replacing it with an ANN. In this case this is an artificial example, but in many observational scenarios and model, one has incomplete models whose deficiencies can be corrected with the NODE approach. In our case replacing the EBM with an ANN is supposed to mirror setups of more realistic models in which one probably has much better knowledge of the governing equations of the atmosphere than the energy balance. In principal it would also be possible to replace the L96 or part of it with an ANN. Setups similar to this, studying the application of NODEs to spatiotemporally chaotic systems have been explored in \citep{npde-pre}.

For the ANN we combine two different kind of layers: (i) dense layers that apply a nonlinear function, called activation function, to a weighted sum of all its inputs, and (ii) convolutional layers that perform a discrete convolution with the convolution filters as learnable parameters. For details on these layers, the reader is referred to standard textbooks such as \citep{Goodfellow-et-al-2016, Alpaydin14}.  

\begin{figure}
    \centering
    \includegraphics[width=0.4\textwidth]{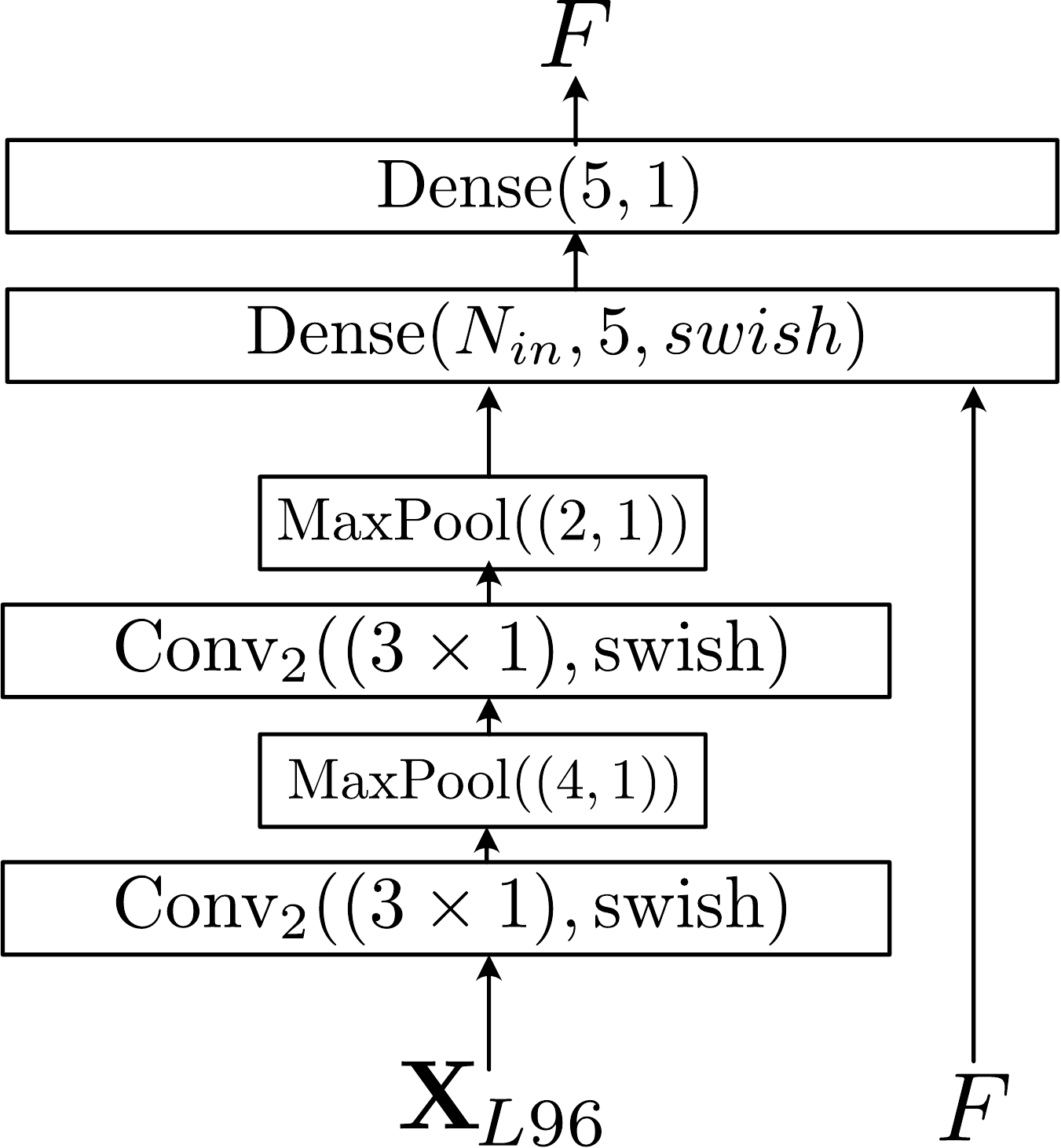}
    \caption{ANN setup used to replace the EBM in the NODE. Convolutional layers with two filters, i.e. channels, a $(3\times1)$ kernel and a swish activation function are used on the L96 variables, the output of these layers and the old forcing value $F$ are used as inputs of two dense layers. $N_{in}$ is chosen to have the correct input dimension which depends on the size of the L96 model.}
    \label{fig:node-setup}
\end{figure}

The ANN $\mathcal{N}$ is set up to have the same input variables as the EBM in Eq.~\ref{eq:model} has arguments: all variables of the L96 model and the EBM itself. Fig.~\ref{fig:node-setup} shows the ANN used. Due to the spatial input, convolutional layers are best suited for those variables and therefore have only the L96 variables as inputs. The forcing, the result of the EBM itself, skips these layers and inputs directly into the dense layers. The swish activation function \citep{ramach2017searching} $\text{swish}(x)=x/(1+\exp{(-x)})$ is used as an activation function and MaxPooling layers reduce the dimension by downsampling it. 

By replacing the EBM the full NODE reads 
\begin{align}
    \dot{F} &= \mathcal{N}(\mathbf{X},F;\Theta)\nonumber\\
    \dot{X}_n &= \left(X_{n+1}-X_{n-2}\right)X_{n-1}-X_n+F;
\end{align}
where $\Theta$ are the parameters of all ANN layers. An adaptive stochastic gradient descent with weight decay (AdamW) \citep{loshchilov2017decoupled} is used to minimize the loss function
\begin{align}
      \mathcal{L}(\Theta) = \sum_{n}\left( X_n - \hat{X}_n(\Theta)\right)^2 + \sum_{i_t, n}\left( F - \hat{F}_n(\Theta)\right)^2 + \gamma \sum_i^{N_\Theta} ||\theta_i||\label{eq:L2}.  
\end{align}
Similar to earlier results for applying NODE techniques to spatiotemporally chaotic systems \citep{npde-pre}, we found that integrating the NODE for long time spans does not significantly decrease the loss on neither the training nor the validation set, but increases the computational complexity massively. Therefore, the NODE is only integrated for $\Delta t=0.05$ with only one time step saved. Hence, we minimize the one-step-ahead loss of the predicted states $(\hat{\mathbf{X}},\hat{F})$ against the training data $(\mathbf{X},F)$. As training data two separated trajectories, each 100 time steps long (at $\Delta t=0.05$), are used. These trajectories are integrated from initial conditions drawn randomly, one from each state within the basins identified by MCBB, i.e. the two shaded area shown in Fig.~\ref{fig:mcbb-1}. The initial $2000$ time steps are discarded to avoid transient dynamics and the following 100 time steps of each of the two trajectories are used as the training set. Subsequent time steps of each of the trajectories are saved as validation set. Thus, the NODE is trained to model the full system with both attractors. 

\begin{figure}
    \centering
    \includegraphics[width=\textwidth]{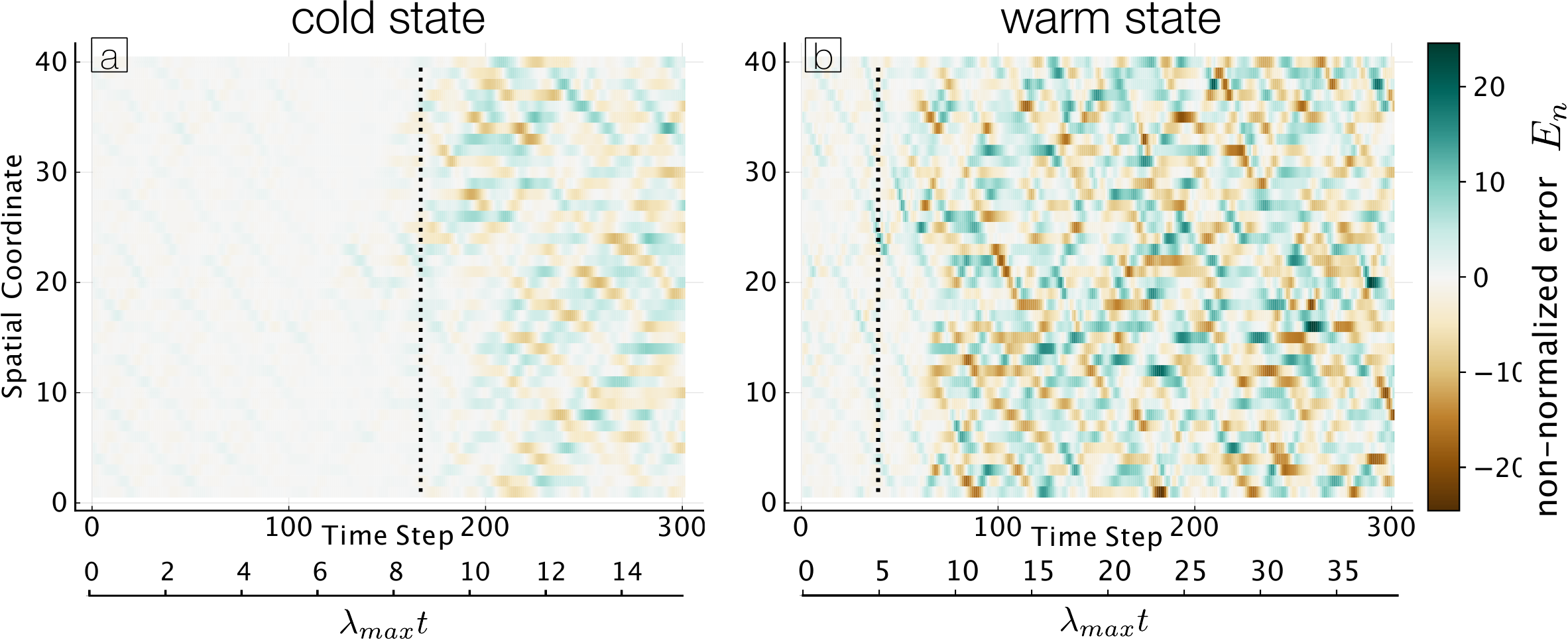}
    \caption{NODE predictions of the Bistable Climate Toy Model, the non-normalized error $E_n(i_t)$ of the L96 variables are shown. (a) shows a prediction on the cold state, (b) the warm state. The valid time $t_v$ is marked with the dashed line.}
    \label{fig:npde-res}
\end{figure}

To evaluate the predictions made by the NODE, we compute a non-normalized error 
\begin{align}
    E_n(i_t) = X_n(i_t) - \hat{X}_n(i_t)
\end{align}
and normalized error 
\begin{align}
    e(i_t) = \frac{||\mathbf{X}(i_t) - \hat{\mathbf{X}}(i_t)||}{<||\mathbf{X}(i_t)||^2>_t^{1/2}}\label{eq:ne}
\end{align}
on the L96 variables, where $<.>_t$ indicates an average over all discrete timesteps $i_t$. The forecast length or valid time of the NODE is then the first time step $t_v$ where $e(i_t) > 0.4$, in accordance with \cite{pathak2018}. The valid time might be expressed in terms of the Lyapunov time $\lambda_{max}t = \lambda_{max}\Delta t\cdot i_t$. Similar to how the NODE was trained with data from both attractors, we also predict and evaluate trajectories from both attractors. Fig.~\ref{fig:npde-res} shows the trajectories of the NODE that were integrated from initial conditions of the first time step outside of the training dataset for both attractors. 

As expected the valid time in terms of discrete time steps is much smaller for the warm state than for the cold state as it is less chaotic, i.e. exhibits a smaller $\lambda_{max}$. In terms of the the rescaled valid time in Lyapunov times it is comparable, yet still a little smaller for the warm state. For the cold state the valid time $169$ time steps or $8.52\lambda_{max}t$ and for the warm state it is $39$ time steps or $5.02\lambda_{max}t$. 

\section{Discussion}

We have presented a framework for addressing the predictability of the first kind and of the second kind in high-dimensional chaotic systems. First, we understand the qualitative properties of the system by discovering the attractors with the largest basins of attractors and evaluate how the volume of these basins changes when control parameters of the system are varied. This might be of great relevance especially in the case several competing asymptotic states, each associated with different basins of attraction, exist. As we gathered knowledge on the attractors of the systems, the NODE approach allows one to predict the evolution of the systems even when the model is only incomplete with respect to the data it is trained with. With the NODE we replaced a sub-module of the model with a data-driven function approximator in the form of an ANN. This approach has the potential to be applied to more complex coupled models in conditions where only incomplete or no knowledge of a specific part of the model is available. When predicting observational data with models, this could also be used to account for unknown or neglected effects in the model. 
The Bistable Climate Toy Model introduced here is an ideal testbed for this approach. The model  is built by coupling a bistable EBM to the L96 model and exhibits two competing attractors in a vast range of the model's parameters. These attractors correspond to a cold and a warm state, for which we are able to identify the basins of attractions and define the bifurcations conducive to tipping points. We are also able to identify accurately which trajectories lead to which of the attractors, so that we use these as training data for the NODE. In the subsequent application of the NODE approach, we purposefully forgot a part of the model, the EBM, and replaced it with an ANN. The NODE can model these introduced deficiencies for both attractors at the same time and make accurate prediction even when only presented with very short training datasets for the data-driven part to be trained on. 
The methods presented are in principle capable of investigating stochastic systems as well which is an exiting avenue of future research with the presented approach. The results on the presented toy model can also be seen as a first step towards analysing and predicting more complex climate models with the presented methods. Especially applying the NODE approach to e.g. atmospheric models and observational data is a highly promising outlook. 

\paragraph*{Acknowledgments} This paper was developed within the scope of the IRTG 1740/TRP 2015/50122-0, funded by the DFG/FAPESP. The authors thank the German Federal Ministry of Education and Research and the Land Brandenburg for supporting this project by providing resources on the high performance computer system at the Potsdam Institute for Climate Impact Research. NB and VL acknowledge funding from the European Union’s Horizon 2020 research and innovation program under grant agreement No 820970 (TiPES). NB acknowledges funding by the Volkswagen foundation. JK acknowledges funding by the Russian Ministry of Education and Science of the Russian Federation Agreement No. 075-15-2020-808. \\

We wish to acknowledge the authors of the Julia libraries DiffEqFlux.jl \cite{rackauckas2020universal}, DifferentialEquations.jl \cite{Rackauskas} and Flux.jl \cite{Flux.jl-2018} that were used for this study.


\begin{thebibliography}{10}
\expandafter\ifx\csname url\endcsname\relax
  \def\url#1{\texttt{#1}}\fi
\expandafter\ifx\csname urlprefix\endcsname\relax\def\urlprefix{URL }\fi
\providecommand{\bibinfo}[2]{#2}
\providecommand{\eprint}[2][]{\url{#2}}

\bibitem{budyko69}
\bibinfo{author}{Budyko, M.~I.}
\newblock \bibinfo{title}{The effect of solar radiation variations on the
  climate of the earth}.
\newblock \emph{\bibinfo{journal}{Tellus}} \textbf{\bibinfo{volume}{21}},
  \bibinfo{pages}{611--619} (\bibinfo{year}{1969}).
\newblock \urlprefix\url{https://doi.org/10.3402/tellusa.v21i5.10109}.
\newblock \eprint{https://doi.org/10.3402/tellusa.v21i5.10109}.

\bibitem{sellers69}
\bibinfo{author}{Sellers, W.~D.}
\newblock \bibinfo{title}{{A Global Climatic Model Based on the Energy Balance
  of the Earth-Atmosphere System}}.
\newblock \emph{\bibinfo{journal}{Journal of Applied Meteorology}}
  \textbf{\bibinfo{volume}{8}}, \bibinfo{pages}{392--400}
  (\bibinfo{year}{1969}).
\newblock
  \urlprefix\url{https://doi.org/10.1175/1520-0450(1969)008<0392:AGCMBO>2.0.CO;2}.
\newblock
  \eprint{https://journals.ametsoc.org/jamc/article-pdf/8/3/392/4975545/1520-0450(1969)008\_0392\_agcmbo\_2\_0\_co\_2.pdf}.

\bibitem{Lucarini_2020}
\bibinfo{author}{Lucarini, V.} \& \bibinfo{author}{B{\'{o}}dai, T.}
\newblock \bibinfo{title}{Global stability properties of the climate:
  Melancholia states, invariant measures, and phase transitions}.
\newblock \emph{\bibinfo{journal}{Nonlinearity}} \textbf{\bibinfo{volume}{33}},
  \bibinfo{pages}{R59--R92} (\bibinfo{year}{2020}).
\newblock \urlprefix\url{https://doi.org/10.1088\%2F1361-6544\%2Fab86cc}.

\bibitem{machowski2008}
\bibinfo{author}{Machowski, J.}, \bibinfo{author}{Bialek, J.} \&
  \bibinfo{author}{Bumby, J.}
\newblock \emph{\bibinfo{title}{Power System Dynamics: Stability and Control,
  2nd Edition}} (\bibinfo{publisher}{Wiley}, \bibinfo{year}{2008}).

\bibitem{Babloyantz1986}
\bibinfo{author}{Babloyantz, A.} \& \bibinfo{author}{Destexhe, A.}
\newblock \bibinfo{title}{Low-dimensional chaos in an instance of epilepsy}.
\newblock \emph{\bibinfo{journal}{Proceedings of the National Academy of
  Sciences}} \textbf{\bibinfo{volume}{83}}, \bibinfo{pages}{3513--3517}
  (\bibinfo{year}{1986}).
\newblock \urlprefix\url{https://www.pnas.org/content/83/10/3513}.
\newblock \eprint{https://www.pnas.org/content/83/10/3513.full.pdf}.

\bibitem{Lytton2008}
\bibinfo{author}{Lytton, W.~W.}
\newblock \bibinfo{title}{Computer modelling of epilepsy}.
\newblock \emph{\bibinfo{journal}{Nature Reviews Neuroscience}}
  \textbf{\bibinfo{volume}{6}} (\bibinfo{year}{2008}).

\bibitem{schwartz2012}
\bibinfo{author}{Schwartz, J.-L.}, \bibinfo{author}{Grimault, N.},
  \bibinfo{author}{Hupé, J.-M.}, \bibinfo{author}{Moore, B. C.~J.} \&
  \bibinfo{author}{Pressnitzer, D.}
\newblock \bibinfo{title}{Multistability in perception: binding sensory
  modalities, an overview}.
\newblock \emph{\bibinfo{journal}{Philosophical Transactions of the Royal
  Society B: Biological Sciences}} \textbf{\bibinfo{volume}{367}},
  \bibinfo{pages}{896--905} (\bibinfo{year}{2012}).
\newblock
  \urlprefix\url{https://royalsocietypublishing.org/doi/abs/10.1098/rstb.2011.0254}.
\newblock
  \eprint{https://royalsocietypublishing.org/doi/pdf/10.1098/rstb.2011.0254}.

\bibitem{Smolen2000}
\bibinfo{author}{Smole, P.}, \bibinfo{author}{Baxter, D.} \&
  \bibinfo{author}{Byrne, J.}
\newblock \bibinfo{title}{Mathematical modeling of gene networks}.
\newblock \emph{\bibinfo{journal}{Neuron}} \textbf{\bibinfo{volume}{26}},
  \bibinfo{pages}{567--580} (\bibinfo{year}{2000}).

\bibitem{Ghil1976}
\bibinfo{author}{Ghil, M.}
\newblock \bibinfo{title}{{Climate stability for a Sellers-type model}}.
\newblock \emph{\bibinfo{journal}{J. Atmos. Sci.}}
  \textbf{\bibinfo{volume}{33}}, \bibinfo{pages}{3--20} (\bibinfo{year}{1976}).

\bibitem{Lucarini2010}
\bibinfo{author}{Lucarini, V.}, \bibinfo{author}{Fraedrich, K.} \&
  \bibinfo{author}{Lunkeit, F.}
\newblock \bibinfo{title}{Thermodynamic analysis of snowball earth hysteresis
  experiment: Efficiency, entropy production, and irreversibility}.
\newblock \emph{\bibinfo{journal}{Q. J. Royal Met. Soc.}}
  \textbf{\bibinfo{volume}{136}}, \bibinfo{pages}{2--11}
  (\bibinfo{year}{2010}).

\bibitem{GhilLucarini2020}
\bibinfo{author}{Ghil, M.} \& \bibinfo{author}{Lucarini, V.}
\newblock \bibinfo{title}{The physics of climate variability and climate
  change}.
\newblock \emph{\bibinfo{journal}{Rev. Mod. Phys.}}
  \textbf{\bibinfo{volume}{92}}, \bibinfo{pages}{035002}
  (\bibinfo{year}{2020}).
\newblock
  \urlprefix\url{https://link.aps.org/doi/10.1103/RevModPhys.92.035002}.

\bibitem{LucariniBodai2017}
\bibinfo{author}{Lucarini, V.} \& \bibinfo{author}{B{\'{o}}dai, T.}
\newblock \bibinfo{title}{Edge states in the climate system: exploring global
  instabilities and critical transitions}.
\newblock \emph{\bibinfo{journal}{Nonlinearity}} \textbf{\bibinfo{volume}{30}},
  \bibinfo{pages}{R32--R66} (\bibinfo{year}{2017}).

\bibitem{LucariniBodai2019PRL}
\bibinfo{author}{Lucarini, V.} \& \bibinfo{author}{B\'odai, T.}
\newblock \bibinfo{title}{Transitions across melancholia states in a climate
  model: Reconciling the deterministic and stochastic points of view}.
\newblock \emph{\bibinfo{journal}{Phys. Rev. Lett.}}
  \textbf{\bibinfo{volume}{122}}, \bibinfo{pages}{158701}
  (\bibinfo{year}{2019}).
\newblock
  \urlprefix\url{https://link.aps.org/doi/10.1103/PhysRevLett.122.158701}.

\bibitem{Lewis2007}
\bibinfo{author}{Lewis, J.~P.}, \bibinfo{author}{Weaver, A.~J.} \&
  \bibinfo{author}{Eby, M.}
\newblock \bibinfo{title}{Snowball versus slushball earth: Dynamic versus
  nondynamic sea ice?}
\newblock \emph{\bibinfo{journal}{Journal of Geophysical Research: Oceans}}
  \textbf{\bibinfo{volume}{112}}, \bibinfo{pages}{C11014}
  (\bibinfo{year}{2007}).
\newblock
  \urlprefix\url{https://agupubs.onlinelibrary.wiley.com/doi/abs/10.1029/2006JC004037}.
\newblock
  \eprint{https://agupubs.onlinelibrary.wiley.com/doi/pdf/10.1029/2006JC004037}.

\bibitem{Abbott2011}
\bibinfo{author}{Abbot, D.~S.}, \bibinfo{author}{Voigt, A.} \&
  \bibinfo{author}{Koll, D.}
\newblock \bibinfo{title}{The jormungand global climate state and implications
  for neoproterozoic glaciations}.
\newblock \emph{\bibinfo{journal}{Journal of Geophysical Research:
  Atmospheres}} \textbf{\bibinfo{volume}{116}} (\bibinfo{year}{2011}).
\newblock
  \urlprefix\url{https://agupubs.onlinelibrary.wiley.com/doi/abs/10.1029/2011JD015927}.
\newblock
  \eprint{https://agupubs.onlinelibrary.wiley.com/doi/pdf/10.1029/2011JD015927}.

\bibitem{Brunetti2019}
\bibinfo{author}{Brunetti, M.}, \bibinfo{author}{Kasparian, J.} \&
  \bibinfo{author}{V{\'e}rard, C.}
\newblock \bibinfo{title}{Co-existing climate attractors in a coupled
  aquaplanet}.
\newblock \emph{\bibinfo{journal}{Climate Dynamics}}
  \textbf{\bibinfo{volume}{53}}, \bibinfo{pages}{6293--6308}
  (\bibinfo{year}{2019}).
\newblock \urlprefix\url{https://doi.org/10.1007/s00382-019-04926-7}.

\bibitem{Margazoglou2020}
\bibinfo{author}{Margazoglou, G.}, \bibinfo{author}{Grafke, T.},
  \bibinfo{author}{Laio, A.} \& \bibinfo{author}{Lucarini, V.}
\newblock \bibinfo{title}{Dynamical landscape and multistability of the earth's
  climate} (\bibinfo{year}{2020}).
\newblock \eprint{2010.10374}.

\bibitem{Lenton1786}
\bibinfo{author}{Lenton, T.~M.} \emph{et~al.}
\newblock \bibinfo{title}{Tipping elements in the earth{\textquoteright}s
  climate system}.
\newblock \emph{\bibinfo{journal}{Proceedings of the National Academy of
  Sciences}} \textbf{\bibinfo{volume}{105}}, \bibinfo{pages}{1786--1793}
  (\bibinfo{year}{2008}).
\newblock \urlprefix\url{https://www.pnas.org/content/105/6/1786}.
\newblock \eprint{https://www.pnas.org/content/105/6/1786.full.pdf}.

\bibitem{lenton2019climate}
\bibinfo{author}{Lenton, T.~M.} \emph{et~al.}
\newblock \bibinfo{title}{Climate tipping points—too risky to bet against}
  (\bibinfo{year}{2019}).

\bibitem{callaway2013dichotomy}
\bibinfo{author}{Callaway, M.}, \bibinfo{author}{Doan, T.~S.},
  \bibinfo{author}{Lamb, J. S.~W.} \& \bibinfo{author}{Rasmussen, M.}
\newblock \bibinfo{title}{The dichotomy spectrum for random dynamical systems
  and pitchfork bifurcations with additive noise} (\bibinfo{year}{2013}).
\newblock \eprint{1310.6166}.

\bibitem{Ashwin2012}
\bibinfo{author}{Ashwin, P.}, \bibinfo{author}{Wieczorek, S.},
  \bibinfo{author}{Vitolo, R.} \& \bibinfo{author}{Cox, P.}
\newblock \bibinfo{title}{Tipping points in open systems: bifurcation,
  noise-induced and rate-dependent examples in the climate system}.
\newblock \emph{\bibinfo{journal}{Philosophical Transactions of the Royal
  Society A: Mathematical, Physical and Engineering Sciences}}
  \textbf{\bibinfo{volume}{370}}, \bibinfo{pages}{1166--1184}
  (\bibinfo{year}{2012}).

\bibitem{Grebogi1983}
\bibinfo{author}{Grebogi, C.}, \bibinfo{author}{Ott, E.} \&
  \bibinfo{author}{Yorke, J.~A.}
\newblock \bibinfo{title}{Fractal basin boundaries, long-lived chaotic
  transients, and unstable-unstable pair bifurcation}.
\newblock \emph{\bibinfo{journal}{Phys. Rev. Lett.}}
  \textbf{\bibinfo{volume}{50}}, \bibinfo{pages}{935--938}
  (\bibinfo{year}{1983}).
\newblock \urlprefix\url{https://link.aps.org/doi/10.1103/PhysRevLett.50.935}.

\bibitem{Vollmer2009}
\bibinfo{author}{Vollmer, J.}, \bibinfo{author}{Schneider, T.~M.} \&
  \bibinfo{author}{Eckhardt, B.}
\newblock \bibinfo{title}{Basin boundary, edge of chaos and edge state in a
  two-dimensional model}.
\newblock \emph{\bibinfo{journal}{New Journal of Physics}}
  \textbf{\bibinfo{volume}{11}}, \bibinfo{pages}{013040}
  (\bibinfo{year}{2009}).
\newblock
  \urlprefix\url{https://doi.org/10.1088\%2F1367-2630\%2F11\%2F1\%2F013040}.

\bibitem{Graham1991}
\bibinfo{author}{Graham, R.}, \bibinfo{author}{Hamm, A.} \&
  \bibinfo{author}{T\'el, T.}
\newblock \bibinfo{title}{Nonequilibrium potentials for dynamical systems with
  fractal attractors or repellers}.
\newblock \emph{\bibinfo{journal}{Phys. Rev. Lett.}}
  \textbf{\bibinfo{volume}{66}}, \bibinfo{pages}{3089--3092}
  (\bibinfo{year}{1991}).
\newblock \urlprefix\url{https://link.aps.org/doi/10.1103/PhysRevLett.66.3089}.

\bibitem{Menck2013}
\bibinfo{author}{Menck, P.~J.}, \bibinfo{author}{Heitzig, J.},
  \bibinfo{author}{Marwan, N.} \& \bibinfo{author}{Kurths, J.}
\newblock \bibinfo{title}{How basin stability complements the linear-stability
  paradigm}.
\newblock \emph{\bibinfo{journal}{Nature Physics}}
  \textbf{\bibinfo{volume}{9}}, \bibinfo{pages}{89--92} (\bibinfo{year}{2013}).
\newblock \urlprefix\url{https://doi.org/10.1038/nphys2516}.

\bibitem{Lorenz1975}
\bibinfo{author}{Lorenz, E.~N.}
\newblock \bibinfo{title}{The physical bases of climate and climate modelling.
  climate predictability}.
\newblock In \emph{\bibinfo{booktitle}{GARP Publication Series}},
  \bibinfo{pages}{132--136} (\bibinfo{publisher}{WMO}, \bibinfo{year}{1975}).

\bibitem{mcbb}
\bibinfo{author}{Maximilian~Gelbrecht, F.~H., Jürgen~Kurths}.
\newblock \bibinfo{title}{Monte carlo basin bifurcation analysis}.
\newblock \emph{\bibinfo{journal}{New Journal of Physics}}
  \textbf{\bibinfo{volume}{22}}, \bibinfo{pages}{033032}
  (\bibinfo{year}{2020}).
\newblock \urlprefix\url{https://doi.org/10.1088\%2F1367-2630\%2Fab7a05}.

\bibitem{pathak2018}
\bibinfo{author}{Pathak, J.} \emph{et~al.}
\newblock \bibinfo{title}{Hybrid forecasting of chaotic processes: Using
  machine learning in conjunction with a knowledge-based model}.
\newblock \emph{\bibinfo{journal}{Chaos: An Interdisciplinary Journal of
  Nonlinear Science}} \textbf{\bibinfo{volume}{28}}, \bibinfo{pages}{041101}
  (\bibinfo{year}{2018}).
\newblock \urlprefix\url{https://doi.org/10.1063/1.5028373}.
\newblock \eprint{https://doi.org/10.1063/1.5028373}.

\bibitem{chen2018neural}
\bibinfo{author}{Chen, R. T.~Q.}, \bibinfo{author}{Rubanova, Y.},
  \bibinfo{author}{Bettencourt, J.} \& \bibinfo{author}{Duvenaud, D.}
\newblock \bibinfo{title}{Neural ordinary differential equations}
  (\bibinfo{year}{2018}).
\newblock \eprint{1806.07366}.

\bibitem{rackauckas2020universal}
\bibinfo{author}{Rackauckas, C.} \emph{et~al.}
\newblock \bibinfo{title}{Universal differential equations for scientific
  machine learning} (\bibinfo{year}{2020}).
\newblock \eprint{2001.04385}.

\bibitem{lorenz96}
\bibinfo{author}{Lorenz, E.}
\newblock \bibinfo{title}{Predictability: a problem partly solved}.
\newblock In \emph{\bibinfo{booktitle}{Seminar on Predictability, 4-8 September
  1995}}, vol.~\bibinfo{volume}{1}, \bibinfo{pages}{1--18}.
  \bibinfo{organization}{ECMWF} (\bibinfo{publisher}{ECMWF},
  \bibinfo{address}{Shinfield Park, Reading}, \bibinfo{year}{1995}).
\newblock \urlprefix\url{https://www.ecmwf.int/node/10829}.

\bibitem{Lorenz2005}
\bibinfo{author}{Lorenz, E.~N.}
\newblock \bibinfo{title}{{Designing Chaotic Models}}.
\newblock \emph{\bibinfo{journal}{Journal of the Atmospheric Sciences}}
  \textbf{\bibinfo{volume}{62}}, \bibinfo{pages}{1574--1587}
  (\bibinfo{year}{2005}).
\newblock \urlprefix\url{https://doi.org/10.1175/JAS3430.1}.
\newblock
  \eprint{https://journals.ametsoc.org/jas/article-pdf/62/5/1574/3483520/jas3430\_1.pdf}.

\bibitem{vanKekem2018PhysD}
\bibinfo{author}{{van Kekem}, D.~L.} \& \bibinfo{author}{Sterk, A.~E.}
\newblock \bibinfo{title}{Travelling waves and their bifurcations in the
  lorenz-96 model}.
\newblock \emph{\bibinfo{journal}{Physica D: Nonlinear Phenomena}}
  \textbf{\bibinfo{volume}{367}}, \bibinfo{pages}{38 -- 60}
  (\bibinfo{year}{2018}).
\newblock
  \urlprefix\url{http://www.sciencedirect.com/science/article/pii/S0167278917301094}.

\bibitem{vanKekem2018NPG}
\bibinfo{author}{van Kekem, D.~L.} \& \bibinfo{author}{Sterk, A.~E.}
\newblock \bibinfo{title}{Wave propagation in the lorenz-96 model}.
\newblock \emph{\bibinfo{journal}{Nonlinear Processes in Geophysics}}
  \textbf{\bibinfo{volume}{25}}, \bibinfo{pages}{301--314}
  (\bibinfo{year}{2018}).
\newblock \urlprefix\url{https://npg.copernicus.org/articles/25/301/2018/}.

\bibitem{Wilks2005}
\bibinfo{author}{Wilks, D.}
\newblock \bibinfo{title}{{Effects of stochastic parametrizations in the Lorenz
  '96 system}}.
\newblock \emph{\bibinfo{journal}{Quarterly Journal of the Royal Meteorological
  Society}} \textbf{\bibinfo{volume}{131}}, \bibinfo{pages}{389--407}
  (\bibinfo{year}{2005}).

\bibitem{Arnold2013}
\bibinfo{author}{Arnold, H.~M.}, \bibinfo{author}{Moroz, I.~M.} \&
  \bibinfo{author}{Palmer, T.~N.}
\newblock \bibinfo{title}{Stochastic parametrizations and model uncertainty in
  the lorenz system}.
\newblock \emph{\bibinfo{journal}{Philosophical Transactions of the Royal
  Society A: Mathematical, Physical and Engineering Sciences}}
  \textbf{\bibinfo{volume}{371}}, \bibinfo{pages}{20110479}
  (\bibinfo{year}{2013}).
\newblock
  \urlprefix\url{https://royalsocietypublishing.org/doi/abs/10.1098/rsta.2011.0479}.
\newblock
  \eprint{https://royalsocietypublishing.org/doi/pdf/10.1098/rsta.2011.0479}.

\bibitem{Vissio2018}
\bibinfo{author}{Vissio, G.} \& \bibinfo{author}{Lucarini, V.}
\newblock \bibinfo{title}{{A proof of concept for scale-adaptive
  parametrizations: the case of the Lorenz '96 model}}.
\newblock \emph{\bibinfo{journal}{Quarterly Journal of the Royal Meteorological
  Society}} \textbf{\bibinfo{volume}{144}}, \bibinfo{pages}{63--75}
  (\bibinfo{year}{2018}).

\bibitem{Chattopadhyay2020}
\bibinfo{author}{Chattopadhyay, A.}, \bibinfo{author}{Hassanzadeh, P.} \&
  \bibinfo{author}{Subramanian, D.}
\newblock \bibinfo{title}{Data-driven predictions of a multiscale lorenz 96
  chaotic system using machine-learning methods: reservoir computing,
  artificial neural network, and long short-term memory network}.
\newblock \emph{\bibinfo{journal}{Nonlinear Processes in Geophysics}}
  \textbf{\bibinfo{volume}{27}}, \bibinfo{pages}{373--389}
  (\bibinfo{year}{2020}).
\newblock \urlprefix\url{https://npg.copernicus.org/articles/27/373/2020/}.

\bibitem{Blender2013}
\bibinfo{author}{Blender, R.} \& \bibinfo{author}{Lucarini, V.}
\newblock \bibinfo{title}{Nambu representation of an extended lorenz model with
  viscous heating}.
\newblock \emph{\bibinfo{journal}{Physica D: Nonlinear Phenomena}}
  \textbf{\bibinfo{volume}{243}}, \bibinfo{pages}{86 -- 91}
  (\bibinfo{year}{2013}).
\newblock
  \urlprefix\url{http://www.sciencedirect.com/science/article/pii/S0167278912002497}.

\bibitem{Sterk2017}
\bibinfo{author}{Sterk, A.~E.} \& \bibinfo{author}{van Kekem, D.~L.}
\newblock \bibinfo{title}{Predictability of extreme waves in the lorenz-96
  model near intermittency and quasi-periodicity}.
\newblock \emph{\bibinfo{journal}{Complexity}} \textbf{\bibinfo{volume}{2017}},
  \bibinfo{pages}{9419024} (\bibinfo{year}{2017}).
\newblock \urlprefix\url{https://doi.org/10.1155/2017/9419024}.

\bibitem{Hu2019}
\bibinfo{author}{Hu, G.}, \bibinfo{author}{B\'odai, T.} \&
  \bibinfo{author}{Lucarini, V.}
\newblock \bibinfo{title}{Effects of stochastic parametrization on extreme
  value statistics}.
\newblock \emph{\bibinfo{journal}{Chaos: An Interdisciplinary Journal of
  Nonlinear Science}} \textbf{\bibinfo{volume}{29}}, \bibinfo{pages}{083102}
  (\bibinfo{year}{2019}).
\newblock \urlprefix\url{https://doi.org/10.1063/1.5095756}.
\newblock \eprint{https://doi.org/10.1063/1.5095756}.

\bibitem{Trevisan2004}
\bibinfo{author}{Trevisan, A.} \& \bibinfo{author}{Uboldi, F.}
\newblock \bibinfo{title}{{Assimilation of Standard and Targeted Observations
  within the Unstable Subspace of the Observation -- Analysis -- Forecast Cycle
  System}}.
\newblock \emph{\bibinfo{journal}{Journal of the Atmospheric Sciences}}
  \textbf{\bibinfo{volume}{61}}, \bibinfo{pages}{103--113}
  (\bibinfo{year}{2004}).

\bibitem{Brajard2020}
\bibinfo{author}{Brajard, J.}, \bibinfo{author}{Carrassi, A.},
  \bibinfo{author}{Bocquet, M.} \& \bibinfo{author}{Bertino, L.}
\newblock \bibinfo{title}{Combining data assimilation and machine learning to
  emulate a dynamical model from sparse and noisy observations: A case study
  with the lorenz 96 model}.
\newblock \emph{\bibinfo{journal}{Journal of Computational Science}}
  \textbf{\bibinfo{volume}{44}}, \bibinfo{pages}{101171}
  (\bibinfo{year}{2020}).
\newblock
  \urlprefix\url{http://www.sciencedirect.com/science/article/pii/S1877750320304725}.

\bibitem{Wilks2006}
\bibinfo{author}{Wilks, D.~S.}
\newblock \bibinfo{title}{Comparison of ensemble-mos methods in the lorenz '96
  setting}.
\newblock \emph{\bibinfo{journal}{Meteorological Applications}}
  \textbf{\bibinfo{volume}{13}}, \bibinfo{pages}{243–256}
  (\bibinfo{year}{2006}).

\bibitem{Duan2016}
\bibinfo{author}{Duan, W.} \& \bibinfo{author}{Huo, Z.}
\newblock \bibinfo{title}{{An Approach to Generating Mutually Independent
  Initial Perturbations for Ensemble Forecasts: Orthogonal Conditional
  Nonlinear Optimal Perturbations}}.
\newblock \emph{\bibinfo{journal}{Journal of the Atmospheric Sciences}}
  \textbf{\bibinfo{volume}{73}}, \bibinfo{pages}{997--1014}
  (\bibinfo{year}{2016}).
\newblock \urlprefix\url{https://doi.org/10.1175/JAS-D-15-0138.1}.
\newblock
  \eprint{https://journals.ametsoc.org/jas/article-pdf/73/3/997/4811703/jas-d-15-0138\_1.pdf}.

\bibitem{Hallerberg2010}
\bibinfo{author}{Hallerberg, S.}, \bibinfo{author}{Paz\'{o}, D.},
  \bibinfo{author}{L\'{o}pez, J.} \& \bibinfo{author}{Rodr\'{\i}guez, M.}
\newblock \bibinfo{title}{{Logarithmic bred vectors in spatiotemporal chaos:
  Structure and growth}}.
\newblock \emph{\bibinfo{journal}{Physical Review E - Statistical, Nonlinear,
  and Soft Matter Physics}} \textbf{\bibinfo{volume}{81}},
  \bibinfo{pages}{1--8} (\bibinfo{year}{2010}).

\bibitem{Carlu2019}
\bibinfo{author}{Carlu, M.}, \bibinfo{author}{Ginelli, F.},
  \bibinfo{author}{Lucarini, V.} \& \bibinfo{author}{Politi, A.}
\newblock \bibinfo{title}{Lyapunov analysis of multiscale dynamics: the slow
  bundle of the two-scale lorenz 96 model}.
\newblock \emph{\bibinfo{journal}{Nonlinear Processes in Geophysics}}
  \textbf{\bibinfo{volume}{26}}, \bibinfo{pages}{73--89}
  (\bibinfo{year}{2019}).
\newblock \urlprefix\url{https://npg.copernicus.org/articles/26/73/2019/}.

\bibitem{AbramovM2008}
\bibinfo{author}{Abramov, R.~V.} \& \bibinfo{author}{Majda, A.~J.}
\newblock \bibinfo{title}{New approximations and tests of linear
  fluctuation-response for chaotic nonlinear forced-dissipative dynamical
  systems}.
\newblock \emph{\bibinfo{journal}{Journal of Nonlinear Science}}
  \textbf{\bibinfo{volume}{18}}, \bibinfo{pages}{303--341}
  (\bibinfo{year}{2008}).
\newblock \urlprefix\url{https://doi.org/10.1007/s00332-007-9011-9}.

\bibitem{Lucarini2011}
\bibinfo{author}{Lucarini, V.} \& \bibinfo{author}{Sarno, S.}
\newblock \bibinfo{title}{{A statistical mechanical approach for the
  computation of the climatic response to general forcings}}.
\newblock \emph{\bibinfo{journal}{Nonlinear Processes in Geophysics}}
  \textbf{\bibinfo{volume}{18}}, \bibinfo{pages}{7--28} (\bibinfo{year}{2011}).

\bibitem{Lucarini2012}
\bibinfo{author}{Lucarini, V.}
\newblock \bibinfo{title}{Stochastic perturbations to dynamical systems: A
  response theory approach}.
\newblock \emph{\bibinfo{journal}{Journal of Statistical Physics}}
  \textbf{\bibinfo{volume}{146}}, \bibinfo{pages}{774--786}
  (\bibinfo{year}{2012}).
\newblock \urlprefix\url{https://doi.org/10.1007/s10955-012-0422-0}.

\bibitem{Gallavotti2014}
\bibinfo{author}{Gallavotti, G.} \& \bibinfo{author}{Lucarini, V.}
\newblock \bibinfo{title}{{Equivalence of Non-equilibrium Ensembles and
  Representation of Friction in Turbulent Flows : The Lorenz 96 Model}}.
\newblock \emph{\bibinfo{journal}{Journal of Statistical Physics}}
  \textbf{\bibinfo{volume}{156}}, \bibinfo{pages}{1027--1065}
  (\bibinfo{year}{2014}).

\bibitem{Vissio2020}
\bibinfo{author}{{Vissio, Gabriele}} \& \bibinfo{author}{{Lucarini, Valerio}}.
\newblock \bibinfo{title}{Mechanics and thermodynamics of a new minimal model
  of the atmosphere}.
\newblock \emph{\bibinfo{journal}{Eur. Phys. J. Plus}}
  \textbf{\bibinfo{volume}{135}}, \bibinfo{pages}{807} (\bibinfo{year}{2020}).
\newblock \urlprefix\url{https://doi.org/10.1140/epjp/s13360-020-00814-w}.

\bibitem{vanKekem2019}
\bibinfo{author}{van Kekem, D.~L.} \& \bibinfo{author}{Sterk, A.~E.}
\newblock \bibinfo{title}{Symmetries in the lorenz-96 model}.
\newblock \emph{\bibinfo{journal}{International Journal of Bifurcation and
  Chaos}} \textbf{\bibinfo{volume}{29}}, \bibinfo{pages}{1950008}
  (\bibinfo{year}{2019}).
\newblock \urlprefix\url{https://doi.org/10.1142/S0218127419500081}.
\newblock \eprint{https://doi.org/10.1142/S0218127419500081}.

\bibitem{Bodai2020}
\bibinfo{author}{B\'odai, T.} \& \bibinfo{author}{Lucarini, V.}
\newblock \bibinfo{title}{Rough basin boundaries in high dimension: Can we
  classify them experimentally?}
\newblock \emph{\bibinfo{journal}{Chaos: An Interdisciplinary Journal of
  Nonlinear Science}} \textbf{\bibinfo{volume}{30}}, \bibinfo{pages}{103105}
  (\bibinfo{year}{2020}).
\newblock \urlprefix\url{https://doi.org/10.1063/5.0002577}.
\newblock \eprint{https://doi.org/10.1063/5.0002577}.

\bibitem{Pierrehumbert2011}
\bibinfo{author}{Pierrehumbert, R.~T.}, \bibinfo{author}{Abbot, D.},
  \bibinfo{author}{Voigt, A.} \& \bibinfo{author}{Koll, D.}
\newblock \bibinfo{title}{Climate of the neoproterozoic}.
\newblock \emph{\bibinfo{journal}{Ann. Rev, Earth Plan. Sci.}}
  \textbf{\bibinfo{volume}{39}}, \bibinfo{pages}{417} (\bibinfo{year}{2011}).

\bibitem{Skufca2006}
\bibinfo{author}{Skufca, J.~D.}, \bibinfo{author}{Yorke, J.~A.} \&
  \bibinfo{author}{Eckhardt, B.}
\newblock \bibinfo{title}{Edge of chaos in a parallel shear flow}.
\newblock \emph{\bibinfo{journal}{Physical Review Letters}}
  \textbf{\bibinfo{volume}{96}}, \bibinfo{pages}{174101}
  (\bibinfo{year}{2006}).

\bibitem{Hirota232}
\bibinfo{author}{Hirota, M.}, \bibinfo{author}{Holmgren, M.},
  \bibinfo{author}{Van~Nes, E.~H.} \& \bibinfo{author}{Scheffer, M.}
\newblock \bibinfo{title}{Global resilience of tropical forest and savanna to
  critical transitions}.
\newblock \emph{\bibinfo{journal}{Science}} \textbf{\bibinfo{volume}{334}},
  \bibinfo{pages}{232--235} (\bibinfo{year}{2011}).
\newblock \urlprefix\url{https://science.sciencemag.org/content/334/6053/232}.
\newblock
  \eprint{https://science.sciencemag.org/content/334/6053/232.full.pdf}.

\bibitem{Ciemer2019}
\bibinfo{author}{Ciemer, C.} \emph{et~al.}
\newblock \bibinfo{title}{Higher resilience to climatic disturbances in
  tropical vegetation exposed to more variable rainfall}.
\newblock \emph{\bibinfo{journal}{Nature Geoscience}}
  \textbf{\bibinfo{volume}{12}}, \bibinfo{pages}{174--179}
  (\bibinfo{year}{2019}).
\newblock \urlprefix\url{https://doi.org/10.1038/s41561-019-0312-z}.

\bibitem{May1977}
\bibinfo{author}{May, R.~M.}
\newblock \bibinfo{title}{Thresholds and breakpoints in ecosystems with a
  multiplicity of stable states}.
\newblock \emph{\bibinfo{journal}{Nature}} \textbf{\bibinfo{volume}{269}},
  \bibinfo{pages}{471--477} (\bibinfo{year}{1977}).
\newblock \urlprefix\url{https://doi.org/10.1038/269471a0}.

\bibitem{Ruelle2009}
\bibinfo{author}{Ruelle, D.}
\newblock \bibinfo{title}{{A review of linear response theory for general
  differentiable dynamical systems}}.
\newblock \emph{\bibinfo{journal}{Nonlinearity}}
  \textbf{\bibinfo{volume}{22(4)}}, \bibinfo{pages}{855--870}
  (\bibinfo{year}{2009}).

\bibitem{Ester1996}
\bibinfo{author}{Ester, M.}, \bibinfo{author}{Xu, X.}, \bibinfo{author}{peter
  Kriegel, H.} \& \bibinfo{author}{Sander, J.}
\newblock \bibinfo{title}{{Density-based algorithm for discovering clusters in
  large spatial databases with noise}}.
\newblock \emph{\bibinfo{journal}{Proceedings Of The Acm Sigkdd International
  Conference On Knowledge Discovery And Data Mining}}
  \textbf{\bibinfo{volume}{pages}}, \bibinfo{pages}{226--231}
  (\bibinfo{year}{1996}).

\bibitem{Berner2017}
\bibinfo{author}{Berner, J.} \emph{et~al.}
\newblock \bibinfo{title}{Stochastic parameterization: Toward a new view of
  weather and climate models}.
\newblock \emph{\bibinfo{journal}{Bulletin of the American Meteorological
  Society}} \textbf{\bibinfo{volume}{98}}, \bibinfo{pages}{565--588}
  (\bibinfo{year}{2017}).
\newblock \urlprefix\url{https://doi.org/10.1175/BAMS-D-15-00268.1}.
\newblock \eprint{https://doi.org/10.1175/BAMS-D-15-00268.1}.

\bibitem{Franzke2015}
\bibinfo{author}{Franzke, C. L.~E.}, \bibinfo{author}{O'Kane, T.~J.},
  \bibinfo{author}{Berner, J.}, \bibinfo{author}{Williams, P.~D.} \&
  \bibinfo{author}{Lucarini, V.}
\newblock \bibinfo{title}{Stochastic climate theory and modeling}.
\newblock \emph{\bibinfo{journal}{Wiley Interdisciplinary Reviews: Climate
  Change}} \textbf{\bibinfo{volume}{6}}, \bibinfo{pages}{63--78}
  (\bibinfo{year}{2015}).

\bibitem{npde-pre}
\bibinfo{author}{Gelbrecht, M.}, \bibinfo{author}{Boers, N.} \&
  \bibinfo{author}{Kurths, J.}
\newblock \bibinfo{title}{Neural partial differential equations for chaotic
  systems}.
\newblock \emph{\bibinfo{journal}{New Journal of Physics}}
  \textbf{\bibinfo{volume}{23}} (\bibinfo{year}{2021}).

\bibitem{loshchilov2017decoupled}
\bibinfo{author}{Loshchilov, I.} \& \bibinfo{author}{Hutter, F.}
\newblock \bibinfo{title}{Decoupled weight decay regularization}
  (\bibinfo{year}{2017}).
\newblock \eprint{1711.05101}.

\bibitem{Lucarini2017}
\bibinfo{author}{Lucarini, V.} \& \bibinfo{author}{Wouters, J.}
\newblock \bibinfo{title}{{Response formulae for n-point correlations in
  statistical mechanical systems and application to a problem of coarse
  graining}}.
\newblock \emph{\bibinfo{journal}{Journal of Physics A: Mathematical and
  Theoretical}} \textbf{\bibinfo{volume}{50}}, \bibinfo{pages}{355003}
  (\bibinfo{year}{2017}).

\bibitem{benettin1980}
\bibinfo{author}{Benettin, G.}, \bibinfo{author}{Galgani, L.},
  \bibinfo{author}{Giorgilli, A.} \& \bibinfo{author}{Strelcyn, J.-M.}
\newblock \bibinfo{title}{Lyapunov characteristic exponents for smooth
  dynamical systems and for hamiltonian systems; a method for computing all of
  them. part 2: Numerical application}.
\newblock \emph{\bibinfo{journal}{Meccanica}} \textbf{\bibinfo{volume}{15}},
  \bibinfo{pages}{21--30} (\bibinfo{year}{1980}).
\newblock \urlprefix\url{https://doi.org/10.1007/BF02128237}.

\bibitem{Datseris2018}
\bibinfo{author}{Datseris, G.}
\newblock \bibinfo{title}{Dynamicalsystems.jl: A julia software library for
  chaos and nonlinear dynamics}.
\newblock \emph{\bibinfo{journal}{Journal of Open Source Software}}
  \textbf{\bibinfo{volume}{3}}, \bibinfo{pages}{598} (\bibinfo{year}{2018}).
\newblock \urlprefix\url{https://doi.org/10.21105/joss.00598}.

\bibitem{Goodfellow-et-al-2016}
\bibinfo{author}{Goodfellow, I.}, \bibinfo{author}{Bengio, Y.} \&
  \bibinfo{author}{Courville, A.}
\newblock \emph{\bibinfo{title}{Deep Learning}} (\bibinfo{publisher}{MIT
  Press}, \bibinfo{year}{2016}).
\newblock \bibinfo{note}{\url{http://www.deeplearningbook.org}}.

\bibitem{Alpaydin14}
\bibinfo{author}{Alpaydin, E.}
\newblock \emph{\bibinfo{title}{Introduction to Machine Learning}}.
\newblock Adaptive Computation and Machine Learning (\bibinfo{publisher}{MIT
  Press}, \bibinfo{address}{Cambridge, MA}, \bibinfo{year}{2014}),
  \bibinfo{edition}{3} edn.

\bibitem{ramach2017searching}
\bibinfo{author}{Ramachandran, P.}, \bibinfo{author}{Zoph, B.} \&
  \bibinfo{author}{Le, Q.~V.}
\newblock \bibinfo{title}{Searching for activation functions}
  (\bibinfo{year}{2017}).
\newblock \eprint{1710.05941}.

\bibitem{Rackauskas}
\bibinfo{author}{Rackauskas, C.} \& \bibinfo{author}{Nie, Q.}
\newblock \bibinfo{title}{Differentialequations.jl – a performant and
  feature-rich ecosystem for solving differential equations in julia}.
\newblock \emph{\bibinfo{journal}{Journal of Open Research Software}}
  \textbf{\bibinfo{volume}{5(1):15}} (\bibinfo{year}{2017}).

\bibitem{Flux.jl-2018}
\bibinfo{author}{Innes, M.} \emph{et~al.}
\newblock \bibinfo{title}{Fashionable modelling with flux}.
\newblock \emph{\bibinfo{journal}{CoRR}}
  \textbf{\bibinfo{volume}{abs/1811.01457}} (\bibinfo{year}{2018}).
\newblock \urlprefix\url{https://arxiv.org/abs/1811.01457}.
\newblock \eprint{1811.01457}.

\end{thebibliography}
\providecommand{\noopsort}[1]{}\providecommand{\singleletter}[1]{#1}%

\end{document}